\documentclass{ifacconf}
\usepackage[
  a4paper,
  left=25mm,
  right=25mm,
  top=25mm,
  bottom=5mm
]{geometry}
\usepackage{lmodern}
\usepackage{amsmath,amssymb}
\usepackage{graphicx}     
\usepackage{natbib}
\usepackage{graphicx}
\usepackage{amsfonts} 
\usepackage{amssymb}
\usepackage{amsmath}
\usepackage{enumerate}
\usepackage{booktabs}

\usepackage{caption}
\usepackage{lastpage}
\usepackage{placeins}
\usepackage{fancyhdr}
\usepackage{amsmath}
\usepackage{bm}

\usepackage{afterpage}
\usepackage{tikz}
\usepackage{listings} 
\usepackage{xcolor}
\makeatletter
\newcommand{\corrauthorfootnote}[1]{%
  \begingroup
  \renewcommand\thefootnote{}%
  \footnotetext{Corresponding author: #1.}%
  \addtocounter{footnote}{-1}%
  \endgroup
}
\makeatother
\allowdisplaybreaks

\begin{document}
\begin{center}
  {\bfseries\Large
  Necessary and Sufficient PID Gain Regions for Global Stabilization of Uncertain Second-Order MIMO Nonlinear Systems\par}

  \vspace{1ex}
\corrauthorfootnote{Cheng Zhao}
  {\normalsize
  Tianyou Xiang and Cheng Zhao*\par}

  \vspace{1ex}

  {\small
  *State Key Laboratory of Mathematical Sciences, AMSS, Chinese Academy of Sciences, Beijing 100190, China.\\
  School of Mathematical Sciences, University of Chinese Academy of Sciences, Beijing 100049, China.\\
  (e-mail: xiangtian0402@163.com, zhaocheng@amss.ac.cn)\par}

  \vspace{1ex}

\end{center}

\vspace{1ex}

\noindent\textbf{Abstract:}\quad
As is well known, classical PID control is ubiquitous in industrial processes, yet a rigorous and explicit design theory for nonlinear uncertain MIMO second-order systems remains underdeveloped. In this paper we consider a class of such systems with both uncertain dynamics and an unknown but strictly positive input gain, where the nonlinear uncertainty is  characterized  by bounds on the Jacobian with respect to the state variables. We explicitly construct a three-dimensional region for the PID gains that is sufficient to guarantee global stability and asymptotic tracking of constant references for all nonlinearities satisfying these Jacobian bounds. We then derive a corresponding necessary region, thereby revealing the inherent conservatism required to cope with worst-case uncertainties. Moreover, under additional structural assumptions on the nonlinearities, these sufficient and necessary regions coincide, yielding a precise necessary-and-sufficient characterization of all globally stabilizing PID gains. All these regions are given in closed form and depend only on the prescribed Jacobian bounds and the known lower bound of the input gain, in contrast to many qualitative tuning methods in the literature.

\vspace{0.7ex}


\section{Introduction}
Despite substantial advances in modern control theory since the 1960s, proportional-integral-derivative (PID) control remains the most widely used feedback strategy in engineering practice. Industrial surveys consistently report that over 95\% of process loops employ PID controllers, with the vast majority configured as PI controllers \cite{aastrom2006advanced}. This enduring dominance stems from their simple structure and satisfactory performance across various applications \cite{ang2005pid}, particularly in motion and process control contexts—such as servo drives and robotics—where the underlying dynamics are often modeled as second-order systems derived from Newton’s second law. For such systems, proportional, integral, and derivative actions naturally compensate for instantaneous errors, steady-state offsets, and trends, respectively \cite{ogata2010modern}, rendering PID a routine practical choice that performs effectively even with limited model information.

In practice, however, one of the main challenges is the systematic and reliable selection of the three PID gains. Extensive research has been devoted to this tuning problem, leading to a wide range of methods based on linear approximations, frequency-response characteristics, or experimentally identified process features, see \cite{ziegler1942optimum}; \cite{cohen1953theoretical}; \cite{ho2003pid}; \cite{aastrom2006advanced}. Collectively, these studies offer a rich toolkit of design techniques applicable to a wide range of conventional control scenarios.

Yet, although most approaches based on linear or locally linearized models can improve control performance without explicit model knowledge, they generally lack rigorous theoretical guarantees for closed-loop stability in strongly nonlinear and uncertain systems (\cite{lei2020feedback}). In a recent study \cite{zhao2017pid}, it was  demonstrated that classical PID control can ensure global stability for a basic class of single-input single-output (SISO) nonlinear uncertain systems with no uncertainty in the control channel, provided that the PID parameters are selected within a three-dimensional unbounded stability region. 

However, most physical systems are inherently multi-input multi-output (MIMO) and are subject to both dynamic uncertainties and input-channel uncertainties. Whether analogous global stability guarantees and explicit gain characterizations can be achieved for MIMO systems remains a challenging issue due to the complex nonlinear dynamics and strong state coupling. Although PID control for MIMO systems has been extensively studied, existing parameter-design results are mostly qualitative or yield only relatively conservative sufficient conditions, see e.g., \cite{alvarez2000pid}; \cite{zhang2019theory}; \cite{cheng2018pid}; \cite{zhao2022towards}. In contrast, establishing both sufficient and necessary conditions for stabilizing PID gains is of fundamental importance, not only from a theoretical perspective but also for practical tuning and implementation. These considerations motivate the present work.

 Motivated by \cite{zhao2017pid} and \cite{zhao2025beyond},  in this paper we explore the capabilities and limits of classical PID control for a class of second-order MIMO nonlinear systems with both dynamic and input-gain uncertainties.  Our main contributions are threefold. First, we explicitly construct a three-dimensional set of PID parameters and demonstate that, for any PID gains within this set, the closed-loop system is globally stable with asymptotic tracking, provided the Jacobian bounds of the uncertainty are known a priori. Second, we derive a necessary parameter region for global stability, revealing an intrinsic gap between necessity and sufficiency inherent to the general uncertain nonlinear structure. Finally, we show that under a strengthened structural assumption, this necessary region also becomes sufficient.

The rest of the paper is organized as follows. The problem formulation will be introduced in the next section. Section~3 presents the main results, while Section~4 contains the proofs. Finally, Section~5 concludes the paper with some remarks on possible extensions. Several auxiliary results are provided in Appendix A.

\section{Problem Formulation}

\subsection{Notation}

For a vector $x\in \mathbb{R}^n$, $|x|$ denotes its Euclidean norm, and $x^\top$ denotes its transpose. For a matrix $M \in \mathbb{R}^{m\times n}$,  the induced norm is defined by $\|M\|=\sup_{|x|=1}|Mx|$. For a square matrix $M\in \mathbb{R}^{n\times n}$, we write $M^{\mathrm{sym}}=(M+M^\top)/2$ for its symmetric part.  For symmetric matrices $M_1$ and $M_2$ of the same order, $M_1\succeq M_2$ ($M_1\succ M_2$) indicates that $M_1-M_2$ is positive semidefinite (definite). Denote by \(C^1(\mathbb{R}^n\times\mathbb{R}^n,\mathbb{R}^n)\) the space of functions \(f:\mathbb{R}^n\times\mathbb{R}^n\to\mathbb{R}^n\) that are continuously differentiable with respect to $x_1$ and $x_2$. For $f=(f_1,f_2,\cdots,f_n)^{\top}\in C^1(\mathbb{R}^n\times \mathbb{R}^n,\mathbb{R}^n)$, the Jacobian matrix with respect to $x_1$ is defined by $
    \frac{\partial f}{\partial x_1}(x_1,x_2)
=[\tfrac{\partial f_i}{\partial x_{1,j}}(x_1,x_2)]_{i,j=1}^n\in\mathbb{R}^{n\times n}$,
and similarly for $\frac{\partial f}{\partial x_2}(x_1,x_2)$. If $f$ is twice continuously differentiable with respect to $x_1$ and $x_2$, then for each component $f_i$ 
we denote its Hessian with respect to $x_1$ by
$\nabla_{x_1}^2 f_i(x_1,x_2)
=[\tfrac{\partial^2 f_i}{\partial x_{1,j}\,\partial x_{1,k}}(x_1,x_2)]_{j,k=1}^n$.
When no confusion arises, we simply write $\nabla_{x_1}^2 f_i$. The notation $\nabla_{x_2}^2 f_i$ is understood analogously.
For a scalar function $U\in C^2(\mathbb{R}^n,\mathbb{R})$, $\nabla U$ denotes its gradient (viewed as a column vector), and $\nabla^2 U$ denotes its  
 Hessian matrix. A smooth vector field $F:\mathbb{R}^n\to \mathbb{R}^n$ is called conservative if it can be written as the gradient of some scalar potential function, i.e. $F(x)=\nabla U(x)$ for some $U:\mathbb{R}^n\to \mathbb{R}$.

\subsection{The control system}
We consider a class of controlled point-mass systems evolving in $\mathbb{R}^n$. Let $p(t)\in\mathbb{R}^n$ denote the position at time $t$, with velocity $v(t)=\dot p(t)$ and acceleration $a(t)=\ddot p(t)$. The external action on the system consists of two parts: a nonlinear term $f(p,v)\in\mathbb{R}^n$ that depends on the state, and a control input $u(t)\in\mathbb{R}^n$ to be designed. The control input acts on the dynamics through an unknown scalar gain $b$, which is assumed to satisfy $b\ge \underline{b}>0$ for a known constant $\underline{b}$. Under the unit-mass normalization, the system dynamics can be written as
\begin{equation}\label{eq:nl-second-order}
  \ddot{{p}}( t ) ={f}( {p}(t ) ,\dot{{p}}(t ) ) +b{u}( t ).
\end{equation}

In this paper, we employ an output-feedback control law with the classical PID structure. Let the tracking error be defined by ${e}(t)= {y}^* - {p}(t)$, where $y^*\in\mathbb{R}^n$ is a constant reference to be tracked. The PID control law is given by
\begin{equation}\label{eq:pid}
  {u}(t) =
  k_p{e}(t) +
  k_i \textstyle{\int}_{0}^{t} {e}(s)\,ds +
  k_d\dot{{e}}(t),
\end{equation}
where $k_p,k_i,k_d\in\mathbb{R}$ are controller gains to be designed. Introducing the state variables
${x}_1( t ) ={p}( t ), \ {x}_2( t) =\dot{{p}}( t )$,
and substituting the PID control law \eqref{eq:pid} into the nonlinear dynamics \eqref{eq:nl-second-order}, the PID controlled MIMO nonlinear system can be written in the state–space form as
\begin{equation}\label{eq:cl-first-order}
\begin{cases}
	\dot{{x}}_1( t ) ={x}_2( t ) ,\\
	\dot{{x}}_2( t ) =f(x_1(t) ,{x}_2( t )) +b{u}( t ) ,\\
	{u}( t ) =k_p {e}( t ) +k_i\textstyle{\int}_0^t{{e}( s) \mathrm{d}s}+k_d \dot{{e}}(t), \\ e(t)=y^*-x_1(t)
\end{cases}
\end{equation}
where ${x}_1(0),{x}_2(0)\in\mathbb{R}^n$ are the initial position and velocity respectively, $y^*\in\mathbb{R}^n$ is the setpoint. The structure in \eqref{eq:cl-first-order} captures a broad class of engineering systems. For example, it encompasses multi-degree-of-freedom mass–spring–damper systems in $\mathbb{R}^n$, where $x_1$ and $x_2$ represent the displacements and velocities of the masses. In this setting, the nonlinear term $f(x_1,x_2)$ models effects such as nonlinear damping, friction, and external disturbances, whereas the control input $u(t)$ represents the generalized forces applied to each degree of freedom.

 The control objective is to design the PID gains such that, under the control law \eqref{eq:pid}, the closed-loop system ensures that the position vector $x_1(t)$ tracks the desired constant setpoint, while the velocity $x_2(t)$ asymptotically converges to zero, for any prescribed reference $y^*$ and any initial position and velocity.

\section{The Main Results}
We develop a systematic framework for the design of PID controllers for several classes of nonlinear uncertain systems, under some structural assumptions on the unknown nonlinearity.
 For given constants $L_1$ and $L_2\geq 0$, we introduce the function class $\mathcal{F}_{L_1,L_2}\subset C^1(\mathbb{R}^n\times\mathbb{R}^n,\mathbb{R}^n)$, which consists of all functions $f$ satisfying
\begin{subequations}
    \begin{align}
&(\tfrac{\partial f}{\partial x_1})^{\mathrm{sym}}\preceq L_1 I,  \ 
 \|\tfrac{\partial f}{\partial x_2}\|\le L_2,\ (x_1,x_2)\in \mathbb{R}^n\times\mathbb{R}^n, \label{eq:4a} \\
 &\tfrac{\partial f}{\partial x_{1}}(x_1,x_2)|_{x_2=0}=\big[\tfrac{\partial f}{\partial x_{1}}(x_1,x_2)|_{x_2=0}\big]^\top,\ \ x_1\in\mathbb{R}^n, \label{eq:4b}
\end{align}
\end{subequations}
where $\tfrac{\partial f}{\partial x_i}$ ($i=1,2$) denote the $n\times n$ Jacobian matrices of $f$ with respect to $x_i$, and $(\tfrac{\partial f}{\partial x_1})^{\mathrm{sym}}:=\tfrac{1}{2}(\tfrac{\partial f}{\partial x_{1}}+(\tfrac{\partial f}{\partial x_{1}})^\top)$ and $I$ is the $n\times n$ identity matrix.

\noindent\textbf{Remark 1.}
In (\ref{eq:4a}), we impose bounds on the Jacobian matrices, where the constants $L_1$ and $L_2$ quantify the uncertainty. These bounds also admit a clear physical interpretation: $L_1$ and $L_2$ represent upper bounds on the 'anti-stiffness' and 'anti-damping' effects of the nonlinear system, respectively. In contrast, \eqref{eq:4b} requires the vector field $f(\cdot,0)$ to be conservative. By Lemma A1, this guarantees the existence of a scalar potential $U$ such that $\nabla U(x_1)=f(x_1,0)$, which is crucial for the Lyapunov construction in the proof of Theorem 1. This requirement is not restrictive in practice and is satisfied by a broad class of physical and engineering systems, including many nonlinear mechanical and electromechanical systems admitting a classical vector Liénard representation.

Before presenting the main result, we first introduce two explicitly constructed parameter regions for the three PID gains. The first one is a \emph{sufficient region} $ \Omega_{\rm pid}^{(1)}$ for global stability, defined as 
\begin{equation}\label{eq:Omega_pid_n}
\begin{aligned}
    \Omega_{\rm pid}^{(1)}=\big\{(k_p,k_i,k_d):&\ k_p>L_1,\ k_d>L_2,\ k_i>0,\ \\[-1mm]
&(k_p-L_1)(k_d-L_2)>k_i+\bar k\ \big\} \, ,
\end{aligned}
\end{equation}
where $\bar k:=2L_2\sqrt{k_i\,(k_d+L_2)}$. The second one is a \emph{necessary region}  $\Omega_{\rm pid}^{(2)}$ for global stability, given by 
\begin{equation}\label{eq:Omega_pid_prime_n}
\begin{aligned}
\Omega_{\rm pid}^{(2)}= \big\{(k_p,k_i,k_d):k_p>L_1,\ k_d>L_2,\ k_i>0,\\[-1mm] (k_p-L_1)(k_d-L_2)>k_i\big \}.
\end{aligned}
\end{equation} 
\begin{thm}
Consider the PID controlled nonlinear MIMO system \eqref{eq:cl-first-order} with uncertain function $f\in \mathcal{F}_{L_1,L_2}$ and input gain $b \geq\underline{b}>0$. Then:

(i) If the triple $(\underline{b}k_p,\underline{b}k_i,\underline{b}k_d)$ lies in $\Omega_{\rm pid}^{(1)}$, then for any $f\in\mathcal{F}_{L_1,L_2}$, any $b\ge \underline{b}$ and any setpoint $y^*$, the closed-loop system satisfies $\lim_{t\to\infty} x_1(t)=y^*$ and $\lim_{t\to\infty} x_2(t)=0$
for any initial states. 

(ii) Conversely, if for every $f\in\mathcal{F}_{L_1,L_2}$, every $b\ge \underline{b}$ and $y^*\in\mathbb{R}^n$, the PID controlled system \eqref{eq:cl-first-order} satisfies $\lim_{t\to\infty}x_1(t)=y^*$ and $\lim_{t\to\infty}x_2(t)=0$ globally, then the triple $(\underline{ b} k_p,\underline{ b} k_i,\underline{ b} k_d)$ must lies in  $\Omega_{\rm pid}^{(2)}$.

\end{thm}
\noindent\textbf{Remark 2.}
Theorem~1 gives two explicitly constructed PID parameter regions: a sufficient region $\Omega_{\rm pid}^{(1)}$  and a necessary region $\Omega_{\rm pid}^{(2)}$. The gap between them, quantified by $\bar{k}$ in~\eqref{eq:Omega_pid_n}, comes from the need to handle the uncertain nonlinear class $\mathcal{F}_{L_1,L_2}$. In particular, $\Omega_{\rm pid}^{(2)}$ describes a fundamental limit: any PID gains that achieve global stability for all admissible nonlinearities must lie inside this region. By contrast, $\Omega_{\rm pid}^{(1)}$ is an explicitly constructed, easy-to-check subset of the gain space that guarantees global stability for every $f\in\mathcal{F}_{L_1,L_2}$. 

Since Theorem 1 provides only a sufficient region and a necessary region, and a gap remains between them. To close this gap and obtain an exact necessary and sufficient condition, we impose a stronger structural assumption on the nonlinearity.
Specifically, we  introduce the function space $\mathcal{G}_{L_1,L_2}$, which is a subset of $\mathcal{F}_{L_1,L_2}$ and consists of all functions 
$f\in C^2(\mathbb{R}^n\times\mathbb{R}^n,\mathbb{R}^n)$ satisfying the following assumptions:

Assumption 1: $f=(f_1,f_2,\cdots,f_n)^{\top}\in \mathcal{F}_{L_1,L_2}$, and $\nabla^2_{x_2} f_i=0$ for all $x_1,x_2\in\mathbb{R}^n$ and every $i=1,\cdots,n$. 

Assumption 2: There exists a scalar function $S$, such that  $\frac{\partial f}{\partial x_2}(x_1,0)=\nabla^2_{x_1} S(x_1)$ for all $x_1\in\mathbb{R}^n$. 

Assumption 2  ensures that the Jacobian $\frac{\partial f}{\partial x_2}$ at $x_2=0$ is a Hessian matrix field, yet it does not explicitly answer what structural properties on $f$ guarantee the existence of such a function $S$. Let us write $A(x_1):=\tfrac{\partial f}{\partial x_2}(x_1,0)$. We now introduce the following assumption, which is equivalent to Assumption 2.  The equivalence can be established using Lemma A2 in the Appendix.

Assumption $2^\prime$:   $A(x_1)$ satisfies $A(x_1)=A(x_1)^\top$, and the following integrability condition holds:
    \begin{equation}
        \tfrac{\partial A_{ij}}{\partial x_{1,k}}(x_1)=\tfrac{\partial A_{ik}}{\partial x_{1,j}}(x_1), \,\forall \, i,j,k\in\{1,\cdots,n\}.
        \label{eq:Rm3-hessian}
    \end{equation}

\noindent\textbf{Remark 3.} 
Under Assumptions 1-2, it can be shown that for every function $f\in\mathcal G_{L_1,L_2}$, the decomposition $f(x_1,x_2)=f(x_1,0)+\nabla_{x_1}^2S(x_1)x_2$ holds. Moreover, condition (\ref{eq:4b}) implies (by Lemma A1 in the Appendix) the existence of a scalar function $U:\mathbb{R}^n\to \mathbb{R}$ such that
$f(x_1,0)=\nabla U(x_1)$. Consequently,
\[f(x_1,x_2)=\nabla U(x_1)+\nabla^2S(x_1)x_2, \text{ for all } x_1,x_2\in\mathbb{R}^n.\]
It is clear that the class $\mathcal{G}_{L_1,L_2}$ is non-empty. In particular, it includes all linear functions of the form $f(x_1,x_2)=Ax_1+Bx_2+c$, where $A=A^\top$, $B=B^\top$, $\lambda_{\max}(A)\le L_1$, and $\|B\|\le L_2$.

Under this strengthened structural assumption, we can characterize a  necessary and sufficient region of PID gains that guarantees global stability of the closed-loop system. 

\setcounter{thm}{0}   
\begin{prop}\label{prop:nD}
Consider the PID controlled MIMO nonlinear system \eqref{eq:cl-first-order} with $f\in \mathcal{G}_{L_1,L_2}$ and $b\in[ \underline{b},\infty)$. Then for any $f\in \mathcal{G}_{L_1,L_2}$, any $b\ge\underline{b}$ and any $y^\ast\in\mathbb{R}^n$, the closed-loop system satisfies $\lim_{t\to\infty} |x_1(t)-y^*|+| x_2(t)|=0$ globally  if and only if the triple $(\underline{b}k_p,\underline{b}k_i,\underline{b}k_d)\in\Omega_{\rm pid}^{(2)}$. 
\end{prop}

\section{Proofs of the main results}
\begin{pf*}{Proof of Theorem~1}
(i) First, we show that if the triple $(bk_p,bk_i,bk_p)$ belongs to $\Omega_{\rm pid}$, then the closed-loop system \eqref{eq:cl-first-order} satisfies
$\lim_{t\to\infty} x_1(t)=y^*,\lim_{t\to\infty} x_2(t)=0$
for any initial conditions and any setpoint $y^*$. 
For notational simplicity,  let $$(k_1, k_0, k_2):=(b k_p, b k_i, b k_d).$$ 
We introduce the transformed variables
$x( t ) =\int_0^t e( \tau ) \mathrm{d}\tau +f( y^*,0 )/k_0$, $y(t)=e(t)$,
$z(t)=\dot{e}(t)$, where $e(t)=y^*-x_1(t)$, and define the auxiliary function
\[g( y,z ) :=f( y^*,0)-f( y^*-y,-z).\]
Then \eqref{eq:cl-first-order} can be rewritten as
\begin{equation}
    \begin{cases}
	\dot{x}=y,\ \ \
	\dot{y}=z,\\
	\dot{z}=g\left( y,z \right) -k_0x-k_1y-k_2z,\\
\end{cases}
\label{eq:cl1-th1}
\end{equation}
where $x,y,z\in \mathbb{R} ^n$. Note that $g(0,0)=0$, hence $(0,0,0)\in\mathbb{R}^n\times \mathbb{R}^n\times\mathbb{R}^n$ is an equilibrium point of \eqref{eq:cl1-th1}.

We next show that the function $g(y,z)$ admits the decomposition $g( y,z ) =B( y ) y+A( y,z ) z$, where $B( y )$ and $A( y,z )$ are defined by
 \begin{equation}
B\left(y\right)=\textstyle{\int}_0^1{\frac{\partial g}{\partial y}\left( \tau y,0 \right) \mathrm{d}\tau}, \ A\left(y,z\right)=\textstyle{\int}_0^1{\frac{\partial g}{\partial z}\left( y,\tau z \right) \mathrm{d}\tau}.
    \label{A,B}
\end{equation}
To this end, for given $y$, define $h( \tau ) =g( \tau y,0 )$ for $\tau\in[0,1]$.
Then $ h^\prime( \tau ) =( \frac{\partial g}{\partial y}( \tau y,0 ) ) y$, so that
\begin{equation}
\begin{aligned}
           g(y,0)= &\ g( y,0 ) -g( 0,0 )=h(1)-h(0) \\
            =&\ \textstyle{\int}_0^1{( \frac{\partial g}{\partial y}( \tau y,0 ) ) y\mathrm{d}\tau}= B( y ) y.
\end{aligned}
    \end{equation}
Similarly, for given $y$ and $z$, define $k( \tau ) =g( y,\tau z )$ for $\tau\in[0,1]$. Then
it can be derived that $g(y,z)-g(y,0)=k(1)-k(0)=A(y,z)z$ with $A(y,z)$ is defined in \eqref{A,B}. 

 Since $f\in\mathcal{F}_{L_1,L_2}$, it follows that $g\in\mathcal{F}_{L_1,L_2}$, and thus
\begin{equation}
    \left\| A( y,z ) \right\| \le \textstyle{\int}_0^1\big\|{\tfrac{\partial g}{\partial z}\left( y,\tau z \right) \big\| \mathrm{d}\tau}\le L_2.
\end{equation}
From the definition of $B(y)$, together with $\frac{\partial g}{\partial y}( y,0) =\frac{\partial f}{\partial x_1}( y^*-y,0 )$ and the symmetry of $\frac{\partial f}{\partial x_1}(x_1,0)$, we conclude that $B(y)$ is symmetric. Besides, from $\tfrac{1}{2}\big(\tfrac{\partial g}{\partial y}\!+\!\big(\tfrac{\partial g}{\partial y}\big)\!^\top\big)\preceq L_1 I$, we have
\begin{equation}
    B(y)  \preceq \textstyle{\int}_0^1 {L_1 I \mathrm{d}\tau} = L_1 I.
    \label{Eq:Byest}
\end{equation}
By using the decomposition of $g$, \eqref{eq:cl1-th1} can be compactly written as
\begin{equation}
    \left[ \begin{array}{c}
	\dot{x}\\
	\dot{y}\\
	\dot{z}\\
\end{array} \right] =M\left( x,y,z \right) \left[ \begin{array}{c}
	x\\
	y\\
	z\\
\end{array} \right] ,
\label{eq:sys-th1}
\end{equation}
where $M:=M\left( x,y,z \right)$ is defined as
 \begin{equation}
    M =\left[ \begin{matrix}
	0_n&		I&		0_n\\
	0_n&		0_n&		I\\
	-k_0I\ &		-k_1I+B\left( y \right)\ &		-k_2I+A\left( y,z \right)\\
\end{matrix} \right] .
\end{equation}

To construct a Lyapunov function, we introduce some additional notation. Define
$\Phi ( y ):= k_1I-B( y ) $ and
$\phi _0:=\mathrm{inf}_y\lambda _{\min}(\Phi ( y ))$.
From \eqref{Eq:Byest} we obtain \[\phi_0=k_1-\mathrm{inf}_y\lambda _{\max}(B( y ))\ge k_1-L_1.\] Similarly, set $\psi :=(\psi_0+\psi_1)/2$, where \begin{align*}\psi _0&=\mathrm{inf}_{y,z}\ \lambda _{\min}( k_2I-A^{\mathrm{sym}}( y,z ) ), \\\psi _1 &= \mathrm{sup}_{y,z}\  \lambda _{\max}(k_2I -A^{\mathrm{sym}}( y,z )).
\end{align*}
Recall 
$\|A( y,z )\|\le L_2$ for all $y,z$, we have 
\begin{align*}-L_2\le  \lambda _{\min}( A^{\mathrm{sym}}( y,z )) \le \lambda _{\max}( A^{\mathrm{sym}}( y,z )) \le L_2,\end{align*}which  implies $k_2-L_2\le \psi _0\le \psi _1\le k_2+L_2$. 
In addition, since $f\in\mathcal{F}_{L_1,L_2}$, we know that 
\[
\tfrac{\partial f}{\partial x_{1}}(x_1,x_2)|_{x_2=0}=\big[\tfrac{\partial f}{\partial x_{1}}(x_1,x_2)|_{x_2=0}\big]^\top, ~\text{for all } x_1\in\mathbb{R}^n.
\]
By Lemma A1 in the Appendix, there exists a potential function $U:\mathbb{R}^n\to \mathbb{R}$ satisfying $f(x_1,0)=\nabla U(x_1)$. 

Utilizing the potential function $U$, we define 
\begin{equation}
    H( y ) \!:=\!\frac{k_1\!-\!\phi _0}{2}| y | ^2\!-U\left( y^*-y \right) +U\left(y^* \right)-\nabla U\left( y^*  \right)^\top\! y,
    \label{eq:Hth1}
\end{equation}
and construct the following Lyapunov function
\begin{equation}
    V(x,y,z)=\left[ x^\top\, y^\top \,z^\top \right] P \left[ x^\top\,y^\top\,z^\top \right] ^\top+H( y ) , \,
    \label{eq:Vth1}
\end{equation}
where the matrix $P$ is
\begin{equation}
P=\frac{1}{2}\begin{bmatrix}
\mu k_0 & k_0 & 0\\
k_0 & \phi_0+\mu\psi & \mu\\
0 & \mu & 1
\end{bmatrix}\otimes I,
\label{eq:pmat}
\end{equation}
and the constant $\mu>0$ in \eqref{eq:pmat} is chosen as
$\mu=\tfrac{ \phi _0\psi _0+k_0 }{2(\phi _0+L_{2}^{2})}$. Here, $\otimes$ denotes the Kronecker product.
By Lemma~A3 in Appendix, the function $V(x,y,z)$ is positive definite and radially unbounded. Moreover, the gradient of $H$ with respect to $y$ is given by 
\begin{align}&\nabla_y H( y ) =( k_1-\phi _0 ) y+f(y^*-y,0)-f(y^*,0)\nonumber \\=&( k_1-\phi _0 ) y-g( y,0 ) =( k_1-\phi _0 ) y-B(y)y.\label{gra:H}
\end{align}
Using the definitions of matrices $P$ and $M$, the time derivative of $V(x,y,z)$ along \eqref{eq:sys-th1} can be written as
 \begin{align*}
        \dot{V}
            &= \left[ x^\top\,y^\top\,z^\top \right]\left(PM\!+\!M^\top P\right)\left[ x^\top\,y^\top\,z^\top \right]^\top
            \!+\nabla_y H(y)^\top \dot y.
\end{align*}After collecting terms we obtain
\begin{align*}
\dot V=\begin{bmatrix}  y^\top\!\! & z^\top \end{bmatrix}
 \begin{bmatrix}
 Q_{11}
  & \tfrac12N_1 \\
 \tfrac12N_1^{\top} & Q_{22}
\end{bmatrix}
\begin{bmatrix}  y\\ z \end{bmatrix}\!+\![(k_1\!-\!\phi_0)y^\top  \!\!- y^\top B(y)^\top] z ,
\end{align*}
where \begin{align*}&Q_{11}=( k_0-\mu k_1 ) I+\mu B(y), \\&N_1=\mu (\psi- k_2)I+\mu A(y,z)-(k_1-\phi_0)I+B(y)^\top, 
\\&Q_{22}=A^{\mathrm{sym}}(y,z)+( \mu -k_2 ) I.\end{align*} Furthermore, define $Q_{12}=-\tfrac{1}{2}( \mu ( \psi -k_2 ) I+\mu A(y,z) )$, then $\dot V$ admits the compact form
\begin{align*}
\dot{V}= -
\begin{bmatrix}  y^\top\!\! & z^\top \end{bmatrix}\begin{bmatrix}
-Q_{11}
& Q_{12}\\
Q_{12}^{\top}
& -Q_{22}
\end{bmatrix}
\begin{bmatrix} y\\ z \end{bmatrix}:= -
\begin{bmatrix}  y^\top\!\! & z^\top \end{bmatrix}Q
\begin{bmatrix} y\\ z \end{bmatrix}.
\end{align*}
We next show that $Q(y,z)$ is positive definite for all $y,z$. Note that
$  \|(\psi  -k_2)  I+ A(y,z)\| \le  | \psi -k_2 |+\| A(y,z) \| $,
and since $| \psi -k_2 |\le |(\psi _1+\psi _0)/2-k_2|\le L_2$, we obtain
$\| Q_{12} \| \le \mu L_2$. By \eqref{eqB1}–\eqref{eqB3} in Appendix,  we know that
 \begin{align*}
    &\lambda_{\min}(-Q_{11})\!=\! \mu \lambda_{\min}(k_1I-B(y))\!-\!k_0\ge \mu \phi _0\!-\!k_0>0,\\
    &\lambda_{\min}(-Q_{22})\!=\! \lambda _{\min}( k_2I-A^{\mathrm{sym}}(y,z) ) \!-\!\mu \ge \psi _0-\mu  >0,  \\
    &\lambda_{\min}(-Q_{11})\lambda_{\min}(-Q_{22})\geq ( \mu \phi _0\!-\!k_0 ) ( \psi _0-\mu ) >\mu ^2L_{2}^{2}.
\end{align*} Therefore, it follows from (\ref{eqB2}) that 
    $\lambda_{\min}(Q(y,z)) \geq \alpha_\ast>0$ for some positive $\alpha_\ast$.
Let $W(x,y,z)=\alpha_\ast(y^2+z^2)$. Then all the requirements in Theorem A1 (LaSalle-Yoshizawa theorem; see \cite{zhao2017pid}) are satisfied. As a consequence, we have $\lim_{t\to\infty}(y(t),z(t))=(0,0)$, which is equivalent to $\lim_{t\to\infty } |x_1(t)-y^*|+|x_2(t)|=0$.

Finally, to complete the proof, it remains to show that for every uncertain $b\geq \underline b$, 
\begin{align*}
(bk_p,bk_i,bk_d)\in \Omega_{\rm pid}^{(1)} \Longleftrightarrow (\underline{b}k_p,\underline{b}k_i,\underline{b}k_d)\in \Omega_{\rm pid}^{(1)}.
\end{align*}
It suffices to show that  if $(k_1,k_0,k_2)\in \Omega_{\rm pid}^{(1)}$, then for all $\alpha\geq 1$, $\alpha(k_1,k_0,k_2)\in \Omega_{\rm pid}^{(1)}$ also holds. For this, we need to verify that the defining inequalities of $\Omega_{\rm pid}^{(1)}$ remain valid. The first three inequalities are obviously preserved. For the last inequality, define
$\zeta(\alpha)=(\alpha k_1-L_1)(\alpha k_2-L_2)-\alpha k_0- 2L_2\sqrt{\alpha k_0(\alpha k_2+L_2)}$. Since $(k_1,k_0,k_2)\in\Omega_{\rm pid}^{(1)}$, we have $\zeta(1)>0$. Differentiating $\zeta$ with respect to $\alpha$ yields
\[
\zeta'(\alpha)
=2\alpha k_1k_2-(L_1k_2+L_2k_1+k_0)
-L_2\rho_{\alpha},
\]
where $\rho_{\alpha}=\frac{2\alpha k_0k_2+k_0L_2}
{\sqrt{\alpha k_0(\alpha k_2+L_2)}}$. By the gain conditions and $\alpha\ge 1$, we have
\begin{align*}
\zeta'(\alpha)
&\ge \left(\alpha k_1 k_2\!-\!L_1 L_2\right)    \!+\!\left\{( k_1\!-\!L_1)( k_2\!-\!L_2)- k_0\right\}    \!-\!L_2\rho_\alpha\\
&\ge L_2\sqrt{k_0}\left(
2\sqrt{k_2+L_2}
-\frac{2\alpha k_2+L_2}{\sqrt{\alpha(\alpha k_2+L_2)}}
\right)  \\
&\ge L_2\sqrt{k_0}\left(
2\sqrt{k_2+L_2}
-\frac{2(\alpha k_2+L_2)}{\sqrt{\alpha(\alpha k_2+L_2)}}
\right)  \\
&=2L_2\sqrt{k_0}
\left(
\sqrt{k_2+L_2}
-\sqrt{k_2+\frac{L_2}{\alpha}}
\right)
\ge 0 .
\end{align*}

Thus $\zeta(\cdot)$ is nondecreasing in $\alpha$, so $\zeta(\alpha)\ge \zeta(1)>0$.

(ii) We use a contradiction argument to show that, if for any $f \in \mathcal{F}_{L_1,L_2}$, $b\ge\underline{b}$ and any
$y^\ast\in\mathbb{R}^n$, the PID controlled system \eqref{eq:cl-first-order} satisfies $\lim_{t\to\infty} x_1(t)=y^*$ and $\lim_{t\to\infty} x_2(t)=0$, then it must hold that
$(\underline{b}k_p,\underline{b}k_i,\underline{b}k_d)\in\Omega_{\rm pid}^{(2)}$.

First, choose $f(x_1,x_2) = L_1 x_1 + L_2 x_2 + c$ and $b=\underline b$, where $c=(c_1,\cdots,c_n)\in\mathbb{R}^n$ is an arbitrary constant vector. Clearly, $f\in\mathcal{F}_{L_1,L_2}$ for any choice of $c$. Write \[x_1:=(x_{11},\cdots,x_{1n})^\top,\ \ x_2:=(x_{21},\cdots,x_{2n})^\top.\] Under this choice of $f$, the PID controlled system \eqref{eq:cl-first-order} decomposes into $n$ completely decoupled scalar subsystems. Consider the subsystem associated with the first coordinate $x_{11}(t)$ and $x_{21}(t)$, then we have 
\begin{equation*}
\begin{cases}
	\dot{{x}}_{11}( t ) ={x}_{12}( t ) ,\\
	\dot{{x}}_{12}( t ) =L_1 x_{11}(t) +L_2x_{21}(t)+c_1+\underline{b}{u_1}( t ) ,\\
	{u}_1( t ) =k_p {e}_1( t ) +k_i\textstyle{\int}_0^t{{e}_1( s) \mathrm{d}s}+k_d \dot{{e}}_1(t), \\ e_1(t)=y^*_1-x_{11}(t)
\end{cases}
\end{equation*}
with initial state ${x}_{11}(0),{x}_{21}(0)\in\mathbb{R}$, setpoint $y^*_1\in\mathbb{R}$.

For this scalar subsystem, Proposition 1 of
\cite{zhao2017pid} shows that a necessary condition for the PID gains to guarantee convergence for all setpoints $y_1^*$, all constants $c_1\in\mathbb{R}$, and all initial states ${x}_{11}(0),{x}_{21}(0)\in\mathbb{R}$ is $(\underline{b}k_p,\underline{b}k_i,\underline{b}k_d)\in\Omega_{\rm pid}^{(2)}$. Therefore, the same condition is necessary for the MIMO case studied here as well.
\end{pf*}
\begin{pf*}{Proof of Proposition~1}

Sufficiency. We first show  that if the triple $(k_1, k_0, k_2):=(b k_p, b k_i, b k_d) \in \Omega_{\rm pid}^{(2)}$ then $\lim_{t\to\infty} |x_1(t)-y^*|+| x_2(t)|=0$. 
We adopt the same notation as in Theorem 1,
\[x( t ) =\textstyle{\int}_0^t e( \tau ) \mathrm{d}\tau +f( y^*,0 )/k_0, y(t)=e(t),
z(t)=\dot{e}(t),\]  and $g( y,z ) =f( y^*,0)-f( y^*-y,-z).$ Since $f\in\mathcal{F}_{L_1,L_2}$,
it follows from the proof of Theorem~1 that
$g( y,z ) =B( y ) y+A( y,z ) z$, where
$B(y)$ and $A(y,z)$ are given in (\ref{A,B}).  
Moreover, $\| A( y ) \|\le L_2$ and
$\lambda _{\max}(B( y ))\le L_1$.

Since $\nabla_{x_2}^2 f_i\equiv 0$, the term $A( y,z )$ is independent of $z$ and can be written as $A(y)=\frac{\partial g}{\partial z}( y,0 )$. Moreover, by Assumption 2, $\frac{\partial f}{\partial x_2}(x_1,0)=\nabla^2_{x_1} S(x_1)$ for some scalar function $S$. Hence,
\begin{equation}
    A(y)=\tfrac{\partial g}{\partial z}( y,0 )=\tfrac{\partial f}{\partial x_2}(y^*-y,0)=\nabla^2_{x_1} S(y^*-y),
\end{equation}
which implies that $A(y)$ is symmetric. 
Now, the system \eqref{eq:cl-first-order} becomes
 \begin{equation}
     \begin{cases}
 	\dot{x}=y,\ \
 	\dot{y}=z,\\
 	\dot{z}= -k_0x-(k_1I-B(y))y-(k_2I-A(y))z
 \end{cases}
 \label{eq:cl1-prop1}
 \end{equation}

Denote
$\Phi(y)= k_1I-B( y )$, $\Psi(y) =k_2I-A( y )$, and introduce the constants $\phi _0=\mathrm{inf}_y
\,\lambda _{\min}(\Phi(y))$, $\psi = (\psi_0+\psi_1)/2$, where
\[\psi _0=\mathrm{inf}_y\lambda _{\min}( \Psi(y)),\ \ \psi _1 = \mathrm{sup}_y
\,\lambda _{\max}( \Psi(y))\  \]
Then it is easy to obtain
$\phi _0\ge k_1-L_1$ and
$\psi _0\ge k_2-L_2$.
Since $f \in \mathcal{G}_{L_1,L_2} \subseteq \mathcal{F}_{L_1,L_2}$, we define $H(y)$  in the same way as in \eqref{eq:Hth1}, and
introduce an auxiliary function
\begin{equation}\label{H:psi}
    H_{\psi}=\mu (\tfrac{k_2-\psi}{2}|y|^2\! + S(y^\ast\!-y) - S(y^\ast)+  \nabla S(y^\ast\!-y)^\top y).
\end{equation}
Consider the following Lyapunov function $V(x,y,z)$
\begin{equation}
    V=\left[ x^\top y^\top z^\top \right] P \left[ x^\top y^\top z^\top \right] ^\top+H( y ) + H_\psi( y ) ,
    \label{eq:Vprop1}
\end{equation}
where the matrix $P$  is defined as in~ $(\ref{eq:pmat})$
with $\mu =\frac{\phi _0\psi _0+k_0}{2\phi _0}$. 
By Lemma A4 in the Appendix, the function $V(x,y,z)$ is positive definite and radially unbounded.

For notational convenience,  denote\begin{align*}
&Q_{11}^\prime=(k_0-\mu k_1)I + \mu B(y)=-\mu\Phi(y)+k_0I, \\&N_1^\prime=\mu(\psi I-\Psi(y))-(\Phi(y)-\phi_0I), 
\\&Q_{22}^\prime=(\mu-k_2)I+A(y)=-\Psi(y)+\mu I.\end{align*}  The time derivative of $V(x,y,z)$  is
{ \begin{align}
        \dot{V}
            &=\begin{bmatrix}  y^\top \!\!& z^\top \end{bmatrix}
     \begin{bmatrix}
 Q_{11}^{\prime}
  & \tfrac12N_1^{\prime} \\
 \tfrac12(N_1^{\prime})^\top & Q_{22}'
\end{bmatrix}
\begin{bmatrix}  y\\ z \end{bmatrix} \nonumber \\ \nonumber
&+((\Phi (y)-\phi _0I)y)^{\top}z+\mu y^{\top}( \Psi ( y ) -\psi I ) z \,\\ \nonumber
&\le -( \mu \phi_0 -k_0 ) |y|^2-( \psi_0  -\mu  ) |z|^2 .
    \nonumber
\end{align}}Recall that $\phi _0\ge k_1-L_1 >0$ and $\psi _0\ge k_2-L_2>0$, which implies $\phi_0\psi_0\geq (k_1-L_1)(k_2-L_2)>k_0$. By the definition $\mu=\frac{\phi _0\psi _0+k_0}{2\phi _0}$, we then obtain
\begin{equation}
    \mu \phi _0-k_0=\tfrac{\phi _0\psi _0-k_0}{2}>0 ,\, \, \psi _0-\mu =\tfrac{\phi _0\psi _0-k_0}{2\phi _0}>0.
    \label{eq:ineqPrpo1}
\end{equation}
This implies $\dot V\le -\beta^*(|y|^2+|z|^2) $ for some $\beta^*>0$. By LaSalle’s theorem,  we know that the equilibrium $(x,y,z)=(0,0,0)$ of \eqref{eq:cl1-prop1} is globally asymptotically stable.

Finally, to complete the proof of the sufficiency part of Proposition~1, it remains to show that for every $b\geq \underline b$, 
\begin{align*}
(bk_p,bk_i,bk_d)\in \Omega_{\rm pid}^{(2)} \Longleftrightarrow (\underline{b}k_p,\underline{b}k_i,\underline{b}k_d)\in \Omega_{\rm pid}^{(2)}.
\end{align*}
This equivalence follows directly from the definition of $\Omega_{\rm pid}^{(2)}$, and the details are therefore omitted.

Necessity. The necessity follows from arguments identical to those used in the proof of Theorem 1 and is therefore omitted.

\end{pf*}

\section{Conclusion}

In this paper, we develop a mathematical theory of PID control for a class of second-order MIMO nonlinear uncertain systems. In contrast to many existing qualitative design approaches, our results provide an explicit construction of the stabilizing PID gains. Specifically, we derive a robust sufficient region $\Omega_{\rm pid}^{(1)}$, and a necessary region $\Omega_{\rm pid}^{(2)}$ for the three PID gains, both given in closed form and depending only on the prescribed Jacobian bounds and the known lower bound of the input gain. This reveals the intrinsic gap that has, to the best of our knowledge, remained unexplored in the literature. Moreover, we demonstrate that, under a strengthened structural assumption on the nonlinearity, the necessary region $\Omega_{\rm pid}^{(2)}$ also becomes sufficient.  
For future investigation, it would be interesting to relax the structural assumptions employed in this paper, and to further study the effects of measurement noise, discrete-time implementation, and time-varying reference signals.

\appendix
\renewcommand{\thesection}{\Alph{section}} 
\renewcommand{\thethm}{\thesection\arabic{thm}} 
\setcounter{thm}{0}                  

\section{Technical Lemmas}\label{sec:app2}              
\setcounter{thm}{0} 
\begin{lem}

(\cite{khalil2002nonlinear}) Let $F:\mathbb{R}^n\to\mathbb{R}^n$ be a smooth vector field. Then $F(x)=\nabla U(x)$ for some scalar function $U$ if and only if \[\tfrac{\partial F}{\partial x}(x)=[\tfrac{\partial F}{\partial x}(x)]^\top, \text{ for all } x \in \mathbb{R}^n.\]
\end{lem}
\begin{lem}
Let $A:\mathbb{R}^n\to\mathbb{R}^{n\times n}$ be a smooth matrix field. Then $A(x)=\nabla^2S(x)$ for some $S$ and all $x\in\mathbb{R}^n$ if and only if $A(x)=A(x)^\top$ and
     \begin{equation}
       \tfrac{\partial A_{ij}}{\partial x_k}(x)=\tfrac{\partial A_{ik}}{\partial x_j}(x), \,\forall \, i,j,k\in\{1,\cdots,n\},\,x\in \mathbb{R}^n.
         \label{eq:Rm3-hessian}
     \end{equation}
\end{lem}
\begin{pf*}{Proof of Lemma~A2} Necessity is immediate from the symmetry of the Hessian and the commutativity of mixed third-order derivatives. For sufficiency, fix $i$ and define
$v_i(x):=(A_{i1}(x),\dots,A_{in}(x)).$ By \eqref{eq:Rm3-hessian}, the Jacobian of $v_i$ is symmetric. Lemma~A1 therefore yields a scalar function $g_i:\mathbb{R}^n\to\mathbb{R}$ such that $v_i(x)=\nabla g_i(x)$, i.e.,
\begin{equation}
\frac{\partial g_i}{\partial x_j}(x)=A_{ij}(x).
\end{equation}
Define $G(x) := (g_1(x), \dots, g_n(x))^\top$ then its Jacobian is $\nabla G(x)=[\tfrac{\partial g_i}{\partial x_{j}}]_{i,j=1}^n=A(x)$.
Since $A(x)=A(x)^\top$, Lemma~A1 applied to $G$ gives a scalar function $S$ such that $G(x)=\nabla S(x)$. Consequently,
\begin{equation}
    \nabla^2 S(x) = \nabla G(x) = [A_{ij}(x)]_{i,j=1}^n = A(x).
\end{equation}

\end{pf*}


\begin{lem}
     Suppose that $(\underline{b}k_p,\underline{b}k_i,\underline{b}k_d)\in\Omega_{\rm pid}^{(1)}$. Then Lyapunov function $V(x,y,z)$ defined in \eqref{eq:Vth1} is positive definite and radially unbounded.
\end{lem}
\begin{pf*}{Proof of Lemma~A3}
It suffices to show matrix $P$ defined by \eqref{eq:pmat} is positive definite, and  the function $H(y)$ defined by \eqref{eq:Hth1} is nonnegative.

    To prove $P\succ0$, we first verify the following inequalities
\begin{align}
    \psi _0&>\mu ,\label{eqB1} \\ 
    ( \mu \phi _0-k_0 ) ( \psi _0-\mu ) &>\mu ^2L_{2}^{2},\label{eqB2} \\ 
    \mu \phi _0&>k_0.\label{eqB3}
\end{align}
    
From $( k_1-L_1 ) ( k_2-L_2 ) -k_0>2L_2\sqrt{k_0( k_2+L_2 )}$, we have $\phi _0\psi _0-k_0>2L_2\sqrt{k_0( k_2+L_2 )}$. By $0<\psi _0\le L_2+k_2$, we obtain $\phi _0\psi _0-k_0>2L_2\sqrt{k_0\psi _0}$, which is equivalent to  \begin{equation} (\phi _0\psi_0+k_0)^2>4(L_{2}^{2}+\phi_0)k_0\psi _0 . \end{equation}
For \eqref{eqB1}, observe that \begin{equation} 2(\psi _0-\mu)=\tfrac{\phi _0\psi_0-k_0+2L_{2}^{2}\psi _0 }{\phi_0+L_{2}^{2}}>0, \end{equation}
and hence \eqref{eqB1} holds. For \eqref{eqB2}, note that
 \begin{align*}
    \begin{aligned}
        &( \mu \phi _0-k_0 ) ( \psi _0-\mu ) -\mu ^2L_{2}^{2}\\
        =&-\mu ^2( \phi _0+L_{2}^{2} ) +\mu ( \phi _0\psi _0+k_0 ) -k_0\psi _0,
\\
=&\tfrac{ 1 }{4(\phi _0+L_{2}^{2})}\left[( \phi _0\psi _0+k_0 ) ^2-4k_0( L_{2}^{2} +\phi_0)\psi _0\right]>0,
    \end{aligned}
\end{align*}
which gives \eqref{eqB2}. Then \eqref{eqB3} follows from \eqref{eqB1}--\eqref{eqB2}. The positive definiteness of $P$ is ensured by $\mu k_0 > 0$ and the following two inequalities.
\begin{align*}
    &\det \left[ \begin{matrix}
	\mu k_0&		k_0\\
	k_0&		\phi _0+\mu \psi\\
\end{matrix} \right] =\mu k_0(\phi _0+\mu \psi )-k_0^{2} \nonumber\\
&>k_0(\mu \phi _0-k_0+\mu ^2\psi _0)>k_0\mu ^2\psi _0>0, \\
   & \det \left[ \begin{matrix}
	\mu k_0&		k_0&		0\\
	k_0&		\phi _0+\mu \psi&		\mu\\
	0&		\mu&		1\\
\end{matrix} \right] =k_0(\mu \phi _0+\mu ^2\psi -k_0-\mu ^3)\nonumber \\
&>k_0(\mu ^2\psi _0-\mu ^3)=k_0\mu ^2( \psi _0-\mu ) >0.
\end{align*}    
    
 To show that $H(y)\geq 0$, we first compute its gradient.
If follows from (\ref{gra:H}) that \[\nabla_y H( y ) =( k_1 I-B(y))y-\phi_0y =( \Phi ( y ) -\phi _0I ) y.\] Moreover, let $\gamma(\tau)=\tau y$. Since $H(0)=0$, the nonnegativeness of $H(y)$ follows from the fact that
\begin{align}
H(y)=&
H(\gamma(1))-H(\gamma(0))=\textstyle{\int}_{0}^{1}\frac{\mathrm d}{d\tau}H(\gamma(\tau))\mathrm d\tau\nonumber \\
=&\textstyle{\int}_{0}^{1}\nabla_y H(\gamma(\tau))^\top \frac{\mathrm d}{\mathrm d\tau }\gamma(\tau)\mathrm d\tau \nonumber\\
=&\textstyle{\int}_{0}^{1} \tau y^\top \big(\Phi(\tau y)-\phi_0 I\big)^\top y\,\mathrm d\tau\ge 0,\label{H:phi}
\end{align}
where the last inequality holds because $\phi_0$ is defined by $\phi _0:=\mathrm{inf}_y\lambda _{\min}(\Phi ( y ))$.

\end{pf*}

\begin{lem} 
Suppose that $(\underline{b}k_p,\underline{b}k_i,\underline{b}k_d)\in\Omega_{\rm pid}^{(2)}$. For \eqref{eq:Vprop1}, there exists a positive definite quadratic form $W$ such that $V\ge W$.
\end{lem}

\begin{pf*}{Proof of Lemma~A4}
From (\ref{H:psi}), we compute \begin{align*}\nabla_y H_\psi(y)  =&\mu((k_2-\psi)y -\nabla^2S(y^*-y)y)\\
=&\mu((k_2-\psi)y-A(y)y)=
\mu(\Psi ( y )-\psi I)y .\end{align*}
Since $\lambda_{\min}(\Psi(y))\geq \psi_0$, the same argument as in (\ref{H:phi}) yields $H_{\psi}( y ) \ge \mu \frac{\psi _0-\psi}{2}| y| ^2$.  Noting also that $H( y ) \ge 0$, we obtain $V \ge W$, where $W$ is given by
\begin{equation*}
    W= \tfrac{1}{2}\mu k_0| x| ^2+k_0x^{\top}y+\tfrac{1}{2}(\phi _0+\mu \psi _0)| y| ^2+\mu y^{\top}z+\tfrac{1}{2}| z| ^2 .
\end{equation*}
The  positive definiteness of $W$ is guaranteed by the following two inequalities:
\begin{align}
    &(\mu \phi_0-k_0)k_0+\mu^2 \psi_0k_0>0, \\
    &\mu k_0(\phi_0+\mu \psi_0)-k_0^2-\mu^3k_0 \nonumber\\
    &=(\mu\phi_0 - k_0)k_0 + \mu^2k_0(\psi_0 -\mu)>0.
\end{align}
By \eqref{eq:ineqPrpo1}, these inequalities hold immediately.

\end{pf*}


\bibliography{ifacconf}             
                                                   







\end{document}